\documentclass[fleqn,usenatbib]{mnras}

\usepackage{newtxtext,newtxmath}
\usepackage[T1]{fontenc}

\DeclareRobustCommand{\VAN}[3]{#2}
\let\VANthebibliography\thebibliography
\def\thebibliography{\DeclareRobustCommand{\VAN}[3]{##3}\VANthebibliography}

\usepackage[mathscr]{euscript}
\usepackage{mathrsfs}
\usepackage{graphicx}	% Including figure files
\usepackage{amsmath}
\usepackage[dvipsnames]{xcolor}
\usepackage[nice]{nicefrac}
\newcommand{\orcid}[1]{\href{https://orcid.org/#1}{\includegraphics[width=8pt]{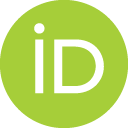}}}
\usepackage{subcaption}
\usepackage{caption}
\usepackage{tabularx}
\usepackage{booktabs}
\usepackage{amsfonts}
\usepackage{mathrsfs}
\usepackage{tikz}
\usepackage{caption}
\usetikzlibrary{arrows.meta, positioning, shadows.blur, backgrounds, fadings, calc, fit}

\definecolor{migRed}{HTML}{D1495B}   % Divergent migration
\definecolor{emriTeal}{HTML}{2A9D8F} % mEMRI
\definecolor{chaosBlue}{HTML}{3A86A8}% Chaotic overlap
\definecolor{axisGray}{HTML}{4A4A4A}
\definecolor{panelGray}{HTML}{F0F0F2}
\definecolor{lineGray}{HTML}{FFFFFF}

\tikzset{
  region/.style={
    rounded corners=12pt,
    align=center,
    text=white,
    blur shadow={shadow blur steps=8, shadow xshift=1.5pt,
                 shadow yshift=-2pt, shadow opacity=35},
  },
  axisarrow/.style={
    -{Stealth[length=5mm, width=3.5mm]},
    line width=2.2pt, color=axisGray!85,
  },
  axislabel/.style={color=axisGray, font=\sffamily\large\bfseries},
}
\title[Lasting mEMRIs]{Lasting mEMRIs: Signatures of Multiple Black Hole Extreme Mass Ratio Inspirals from AGN discs}

\author[Moncrieff et al.]{
Jordan W. N. Moncrieff$^{1,2}$\thanks{E-mail: jordan.moncrieff@research.uwa.edu.au}
\orcid{0009-0009-1028-5381}, 
Evgeni Grishin$^{2,3}$\thanks{E-mail: evgeni.grishin@monash.edu.au} 
\orcid{0000-0001-7113-723X}, %
Fiona H. Panther$^{1,2}$
\orcid{0000-0002-2618-5627},
Alessandro A. Trani$^{4,5}$
\orcid{0000-0001-5371-3432}
\\
$^{1}$ School of Physics, Mathematics and Computing, University of Western Australia, Crawley WA 6009, Australia \\
$^{2}$OzGrav: Australian Research Council Centre of Excellence for Gravitational Wave Discovery, Australia\\
$^{3}$School of Physics and Astronomy, Monash University, Clayton, VIC 3800, Australia\\
$^4$Departamento de Astronom\'ia, Facultad Ciencias F\'isicas y Matem\'aticas, Universidad de Concepci\'on, 4030000, Concepci\'on, Chile\\
$^5$National Institute for Nuclear Physics – INFN, Sezione di Trieste, I-34127, Trieste, Italy\\
}

\date{Accepted XXX. Received YYY; in original form ZZZ}

\pubyear{2026}

\begin{document}
\label{firstpage}
\pagerange{\pageref{firstpage}--\pageref{lastpage}}
\maketitle

\begin{abstract}
Encounters of stellar-mass black holes (sBH) in active galactic nuclei discs naturally form mean-motion resonances of sBH pairs. These  mean-motion resonances can be broken close to the central supermassive black hole when gravitational-wave emission dominates over gas torques. We derive a simple condition that depends only on the sBH mass ratio and the resonance index for which a pair of sBH in a mean-motion resonance will remain intact and can be detectable as a multiple Extreme-Mass-Ratio-Inspirals (mEMRIs) in the LISA band.
We present for the first time the gravitational-wave strain signatures of a pair of co-migrating sBH in mean-motion resonance, including additional gravitational-wave harmonics associated with orbital conjunctions of the individual EMRIs. The frequency evolution and the associated chirp mass is degenerate with isolated EMRIs of different masses. However, the mEMRI mass ratio is quantised by the resonance order and is larger than the true mass ratio. The degeneracy can be broken if the true sBH mass ratio can be inferred from individual strain amplitudes. We argue that mEMRIs should be common if disc driven EMRIs are to be detected by LISA, with $\sim 10$ mEMRIs per year in the local universe ($z \lesssim 0.2$), highlighting the need to model mEMRI systems in future LISA data analysis.
\end{abstract}


\section{INTRODUCTION }
\label{sec:Introduction}

The accretion discs of active galactic nuclei (AGN) are among the most dynamically rich environments in the Universe for compact objects. These include the central supermassive black hole (SMBH) and stellar-mass black holes (sBHs), which are either captured from the surrounding nuclear star cluster \citep{bartos2017rapid,tagawa2020formation} or formed in situ in the disc's self-gravitating outer regions \citep{stone2017assisted,gilbaum2022feedback}. Gas drag drives sBHs onto nearly circular, coplanar orbits \citep{generozov23, wang24}, after which disc torques cause them to migrate inward \citep{bellovary2016migration,grishin2024effect}. The dense gas also leads to sBH--sBH binary formation through dynamical encounters at low relative velocities \citep{row23,moncrieffmerger2025}, making AGN discs a prolific channel for gravitational-wave (GW) mergers detectable by LIGO--Virgo--KAGRA (LVK) \citep{mckernan2020black,tagawa2020formation}.

In the absence of migration traps, whose locations depend on the AGN disc profile and sBH masses, sBHs may migrate close enough to the central SMBH for GW emission to drive a merger within the AGN lifetime \citep{pan2021formation,gilbaum2025escape,sun2025emriform}. The final inspiral produces an extreme mass-ratio inspiral (EMRI), potentially detectable by the Laser Interferometer Space Antenna (\textit{LISA}) \citep{amaro2023astrophysics} or \textit{Tianqin} \citep{TianqinAstro2025}. The imprint of the AGN environment on the EMRI waveform, through gas drag, nearby perturbers, and relativistic precession, offers a unique probe of disc physics and the SMBH spacetime \citep{barausse2007EMRI,kocsis2011emri}. Moreover, combining the GW-inferred luminosity distance with an electromagnetic redshift enables a standard-siren measurement of the Hubble constant \citep{laghi2021cosmology}.

Most studies of BH dynamics in AGN discs incorporate simplified 1D prescriptions for the interaction of sBHs during encounters (\citealp[e.g.,][]{tagawa2020formation, vaccaro2026role}, although see \citealp{rowan26} for a recent hybrid 3D N-body simulations). However, N-body simulations coupled with semi-analytic prescriptions for gas torques show that differential migration naturally leads to the formation of mean motion resonances \citep[MMRs, ][]{secunda19,lott2025scattering,moncrieffmerger2025,Gonglewski2026hierarchicalMMR}. MMRs maintain the period ratio of two orbiting objects at a simple integer ratio, preventing further differential migration, close encounters, and binary formation. These MMRs have been shown to be important in determining the expected population properties of sBH mergers in AGN discs \citep{moncrieffmerger2025,epstein-mmr2025,Gonglewski2026hierarchicalMMR}, and can remain stable until very close to the central SMBH \citep{peng2023synchronizing,peng2024fate}.
A detection of a sBH pair in MMR with  \textit{LISA} and/or \textit{Tianqin} could lead to unique observational signatures.

\begin{figure*}
    \includegraphics[width=\textwidth]{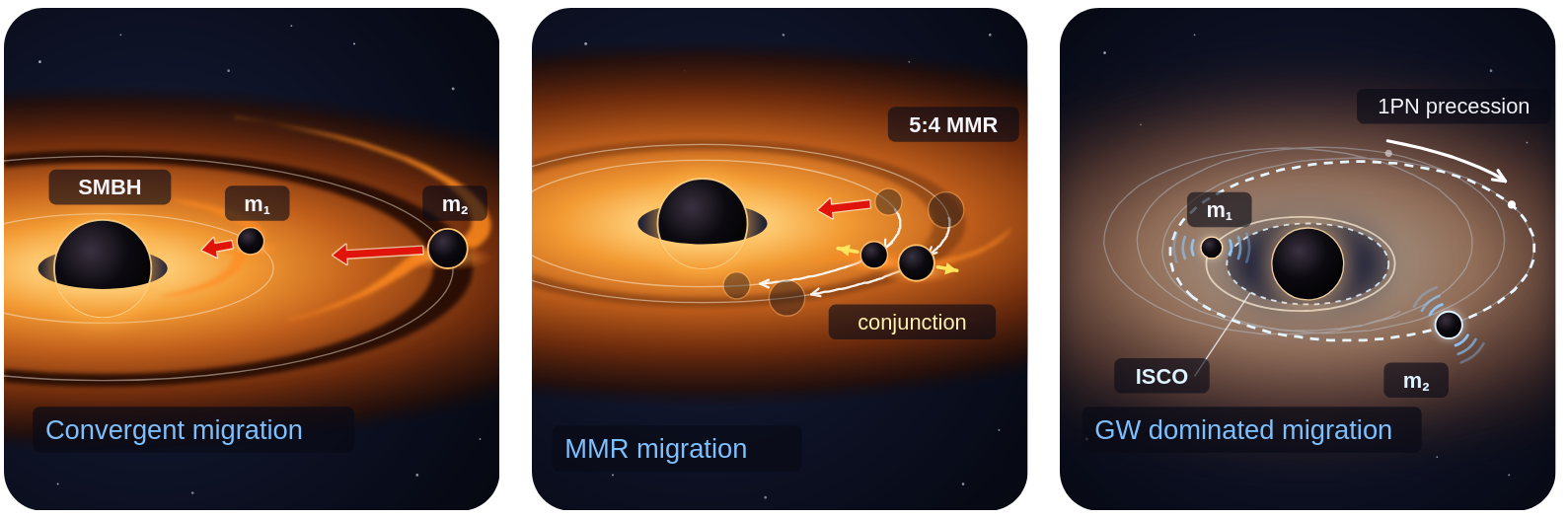}
    \caption{Cartoon illustrating the process of mEMRI formation in AGN disc. Left panel: convergent migration from gas torques brings the two BHs together, with the outer (more massive) massive BH migrating faster under type II migration, until they are captured into MMR. Middle panel: the MMR prevents further convergence, and the BH migrate in lock-step from the gas dominated region closer to the SMBH where GW dissipation is more dominant. Right panel: If $m_1/m_2\lesssim q_{\rm crit}$ (see Eq. \ref{eq:q_crit}), the MMR remains stable in this GW dominated regime, migrating to the highly relativistic regime through GW emission.}
    \label{fig:cartoon}
\end{figure*}

In this paper, we study the processes of MMR formation, co-migration in MMR and multiple EMRIS (mEMRI), sketched in Figure \ref{fig:cartoon}. Initially, two sBHs embedded in an AGN disc experience convergent migration, until they are captured into MMR. They then migrate in lock-step through gas torques towards the central SMBH. Eventually, the torques from GW emission overtake those from interactions with the accretion disc, and the system remains in MMR only if $m_1/m_2$ is below some critical value (see Section \ref{sec:BHsInAGN}). In the relativistic regime, differential apsidal precession drives eccentricity increase of the outer mass, leading to non-negligible eccentricity. If $M_{\rm SMBH}\lesssim 10^6M_{\odot}$, the two masses will enter the LISA frequency band, being detectable as a multiple EMRI.

We begin in Section \ref{sec:BHsInAGN} by overviewing sBH dynamics in AGN discs and our numerical setup. We derive a simple criterion for MMR stability against differential GW-induced migration and verity it with numerical examples of MMR formation, break-up and retention in the relativistic limit. In Section \ref{sec:Relativistic_evo} we study the relativistic effects on the evolution of the stable mEMRI on apsidal anti-alignment, eccentricity growth and offset from exact resonance. In Section \ref{sec:mEMRI_signals} we analyse the GW signal from a fiducial mEMRI system, showing unique signatures. In Section \ref{sec:LISA_discussion} we discuss the impact of mEMRIs on LISA data analysis, degeneracies and characteristics. In Section \ref{sec:astro_implications}, we discuss the the astrophysical implications, rates, comparison to past work, caveats and future prospects. We can expect conservatively $\sim10$ mEMRIs per year in the local universe ($z<0.2$) if AGN disc driven EMRI production is efficient, constituting a significant fraction of the observed EMRI population. Finally, we summarise our main results in Section \ref{sec:summary}.

\section{MMR formation and breakup in AGN discs}\label{sec:BHsInAGN}

We follow our setup in \cite{moncrieffmerger2025}, where the AGN disc model is first created with \texttt{pAGN} \citep{pAGN-code}. We then set up a three body system with the 3.5PN regularised few-body code \texttt{TSUNAMI} \citep{tsunami-code,tra2024}, with a central SMBH and two sBH. Each sBH is assigned a local gas torque and eccentricity damping corresponding to the AGN model. The radial torques include type I \citep{ward1997protoplanet}, type II \citep{kanagawa2018radial}, and thermal \citep{masset2017coorbital, grishin2024effect, gilbaum2025escape} torques. In particular, type II migration can alter the location (or even the existence) of migration traps, where the radial net torque changes sign \citep[see][for details]{moncrieffmerger2025}.

\begin{figure*}
    \centering
    \includegraphics[width=0.91\textwidth]{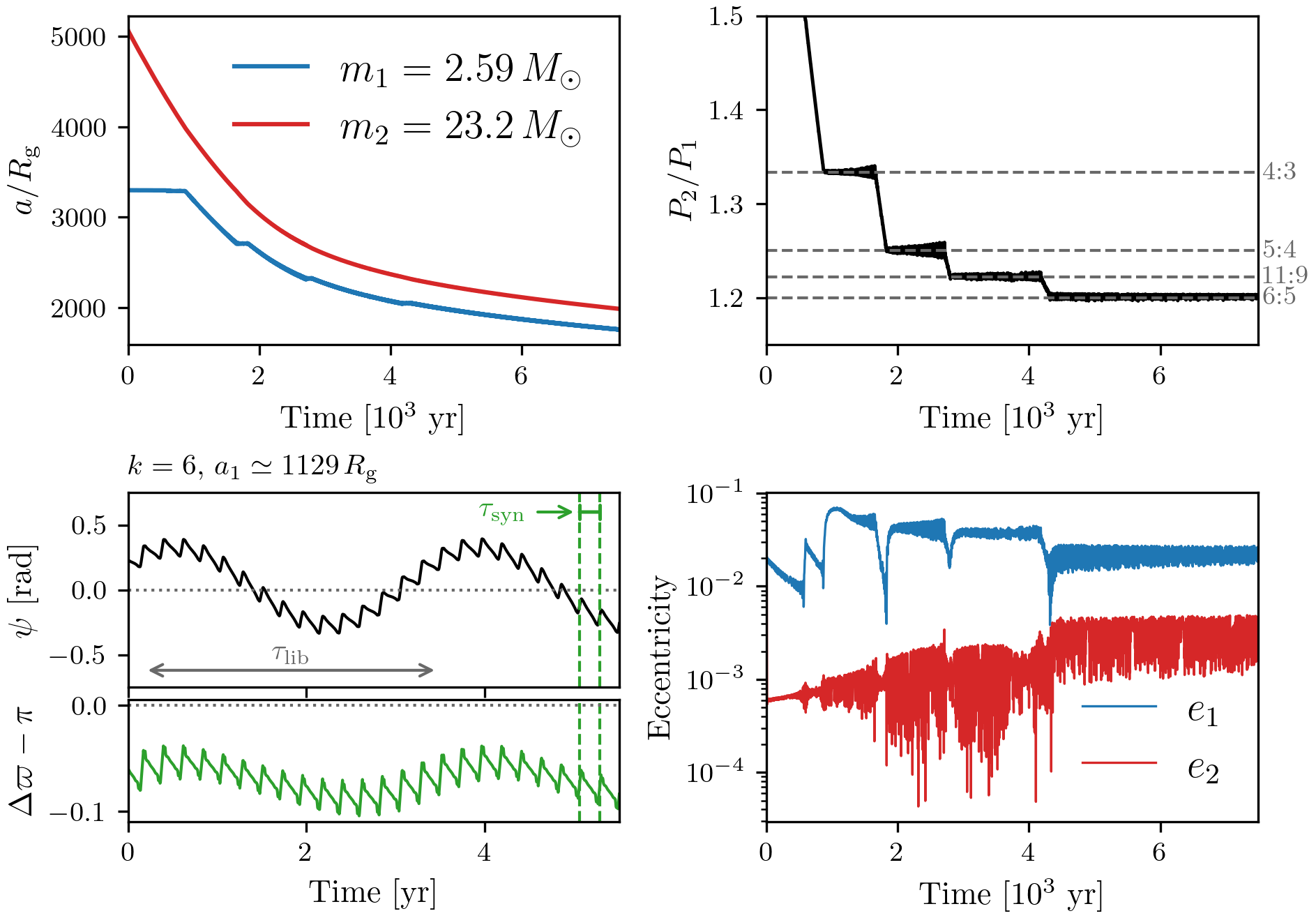}
    \caption{Formation of MMR in the AGN disc. We initialise two sBH of mass $m_1=2.59M_\odot$ at distance $a_1=3300 R_g$ and $m_2=23.2M_\odot$ at $a_2=5060 R_g$ in an AGN disc model of SMBH mass $M_{\rm SMBH} = 1\times10^6 M_\odot$, Eddington ratio $\dot{m}= 1$ and $\alpha$-viscosity $\alpha=0.01$. We see that $m_1$ is initially in a migration trap, until it is captured into a series of MMRs with the inward migrating (gap opening) mass $m_2$, resulting in the two sBHs co-migrating. The bottom left panel shows a zoomed in view of the resonance over six years, including the resonance angle $\psi$ and the  difference in arguments of pericentre $\Delta \varpi$.}
    \label{fig:MMR_formation}
\end{figure*}

\textit{MMR Formation:} As a first example. we set up a Sirko-Goodman AGN disc model \citep{2003MNRAS.341..501S} with $M_{\rm SMBH}=10^6 M_\odot$, accretion rate $\dot{m}\equiv \dot{M}_{\rm SMBH}/\dot M_{\rm Edd}=1$, and $\alpha$-viscosity parameter $\alpha=0.01$. The smaller mass $m_1=2.59M_\odot$ orbits at a stable orbit at a migration trap near $\approx 3300 R_{\rm g}$ ($R_{\rm g}=GM_{\rm SMBH}/c^2$). The heavier sBH of $m_2=23.2M_\odot$ migrates inward. The top left panel of Fig. \ref{fig:MMR_formation} shows the radial decay of both sBH and the capture a MMR. 

MMR are generally quantified by the condition $kP_1 - (k-p)P_2 \approx 0$, where $k$ is the resonance index and $p$ is the resonance order. Usually the first order resonances $p=1$ are the strongest, which is evident in the middle left panel of Fig. \ref{fig:MMR_formation}, where the period ratios move between several first order resonances (and one second order) until converging to the $k=6$ resonance (top right). Capture into MMR can be shown more carefully by considering the resonance angle $\psi$, defined by:
\begin{equation}\label{eq:res_angle}
    \psi=k\lambda_2-(k-1)\lambda_1-\phi_1,
\end{equation}
where $\lambda_i$ is the mean longitude, and $\phi_1$ is an effective two body longitude of pericentre \citep[e.g.,][]{batygin2013res}, given by approximately \citep{deck2015resonance,epstein-mmr2025}:

\begin{equation}
  \phi_1=\arctan\!\left(
    \frac{e_1\sin\varpi_1-e_2\sin\varpi_2}
         {e_1\cos\varpi_1-e_2\cos\varpi_2}
  \right),
  \label{eq:phi1}
\end{equation}
where $\varpi_i$ is the longitude of pericentre, and $e_i$ is the eccentricity of body $i$.

We see the librating resonance angle in the bottom left panel of Fig. \ref{fig:MMR_formation}. Both sBH orbits are anti-aligned, where the difference in their longitude of pericentre $\Delta \varpi \equiv |\varpi_2 - \varpi_1|$ is near $\pi$, which is the most stable configuration.

Besides the orbital periods, the additional characteristic timescales for the MMR are the synodic period and the libration timescale. The synodic period $\tau_{\rm syn} = (k-1)P_2 = kP_1$ is the time between conjunctions and is associated with rapid changes in the orbital elements due to the sBH pair interaction. The time for the resonance angle to librate is the libration timescale $\tau_{\rm lib} \sim P_2 (M_{\rm SMBH}/(m_1+m_2))^{2/3}k^{-4/3}$ \citep{batygin2015capture, moncrieffmerger2025}. The ratio of the two timescales is $\tau_{\rm lib}/\tau_{\rm syn} \sim (M_{\rm SMBH}/(m_1+m_2))^{2/3}k^{-7/3} \approx 17.5$, which is correct to order unity when compared to the simulation in Figure \ref{fig:MMR_formation}.

\begin{figure*}
    \centering
    \includegraphics[width=0.9\textwidth]{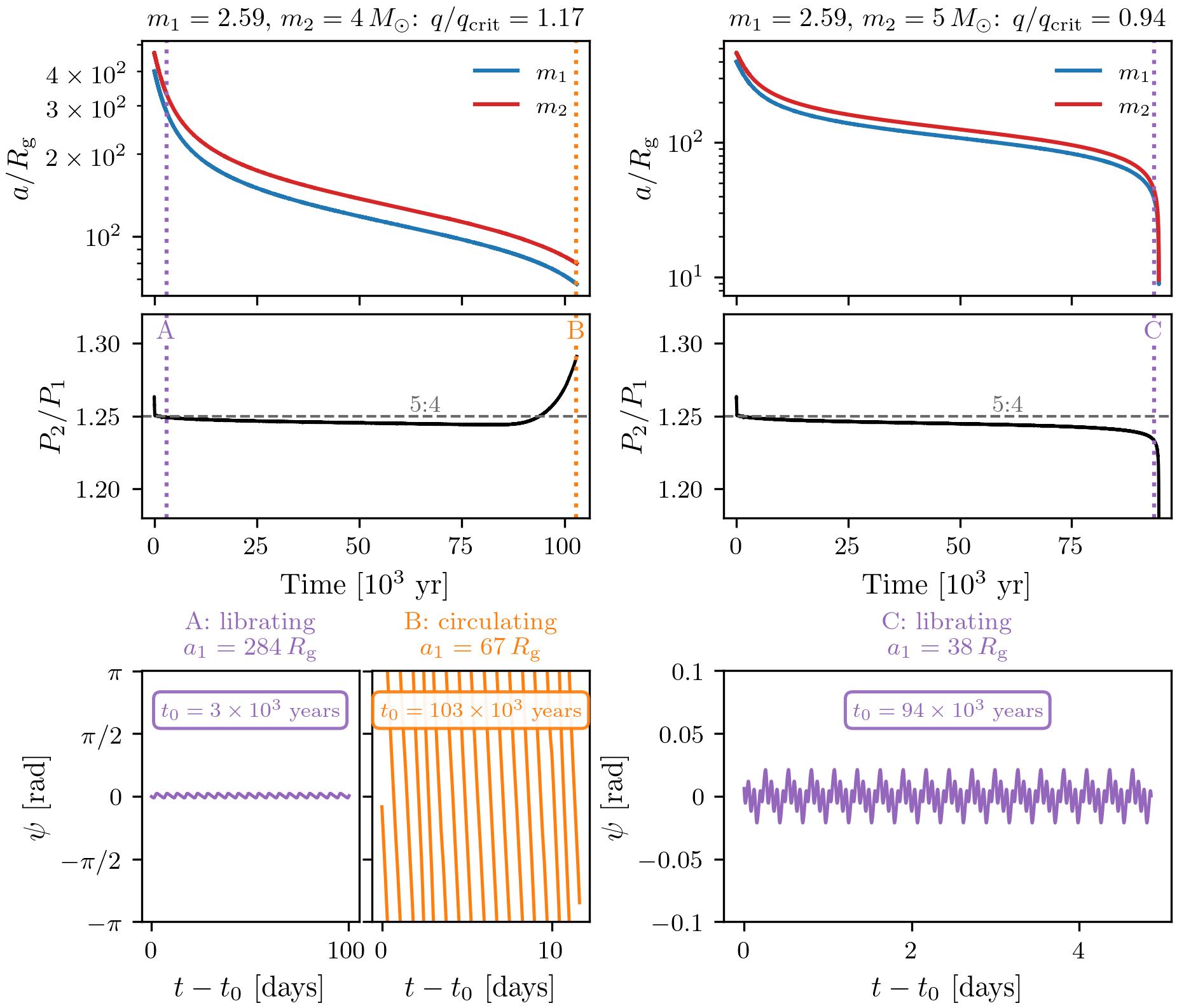}
    \caption{The evolution of two near identical MMR systems into the gravitational wave dominated regime. The left column system has $q>q_{\rm crit}$, the right $q<q_{\rm crit}$. The left system goes from in resonance (librating) to out of resonance (circulating) once GW torques become strong enough (well outside the EMRI band). Meanwhile, the right system remains librating well into the GW dominated regime (mEMRI).}
    \label{fig:div_mig_break}
\end{figure*}

\textit{MMR Breakup:} It was shown in \cite{moncrieffmerger2025} that resonance capture of convergently migrating sBH in AGN discs can be analytically quantified by the adiabatic criteria \citep[e.g.,][]{batygin2015capture}, however, less attention was devoted to MMR breaking. The proximity to the SMBH leads to a transition where the migration is governed by GW dissipation, and gas torques become negligible\footnote{The radial location in the disc where this occurs is dependent on the AGN disc model, but it generally occurs at radii $\sim30-100R_{\rm g}$.}. This leads to a semi-major axis evolution $\dot{a}_{i}=-a_{i}/\tau_{i}$, where the GW timescale is given by \citep{Peters64}:
\begin{equation}
    \tau_{i}=\frac{a_{i}}{\dot{a}_{i}} =\frac{5c^{5}a_{i}^{4}}{64G^{3}M_{\rm SMBH}^2m_{i}}.
\end{equation}

The relative migration rate $\tilde{\tau}=(\tau_2^{-1}-\tau_1^{-1})^{-1}$ is then:
\begin{equation}
    \tilde{\tau}=\frac{5c^{5}}{64G^{3}M_{\rm SMBH}^2}\left( \frac{m_{2}}{a_{2}^{4}}-\frac{m_{1}}{a_{1}^{4}} \right)^{-1}
\end{equation}
Migration will become divergent when $\tilde{\tau}<0$ or when $\tau_1 < \tau_2$, leading to $q=(m_1/m_2)>(a_1/a_2)^4$. For a condition at exact resonance $a_1/a_2=[(k-1)/k]^{2/3}$, divergent migration occurs if
\begin{equation}
    \frac{m_1}{m_2} \gtrsim \left( \frac{k-1}{k} \right)^{8/3}\equiv q_{\rm crit}. \label{eq:q_crit}
\end{equation}

For an illustrative example, we take a disc model of $M_{\rm SMBH}=6\times 10^5 M_\odot$, $\dot{m}=0.5$ and $\alpha=0.01$. We normalise the distances with units of the gravitational radius $R_g \equiv GM_{\rm SMBH}/c^2$. We run two simulations where in both the lighter sBH of $m_1=2.59M_\odot$ is initialised at $400R_{\rm g}$. The more massive sBH is set just outside the $k=5$ resonance at $a_2\approx464R_{\rm g}$. The mass of $m_2$ is either $4M_\odot$ or $5M_\odot$. In this case, $q_{\rm crit}=(4/5)^{8/3}\approx0.55$, so the transitional mass for divergent migration would be $m_{2,\rm{crit}}=2.59M_\odot/q_{\rm crit} \approx 4.7M_\odot$. The simulation with $m_2=4M_\odot<m_{2,\rm{crit}}$ is expected to breakup once the pair enter the GW dominated regime, while the simulation with $m_2=5M_\odot > m_{2,\rm{crit}}$ is expected to remain in resonance until very close to the SMBH.

Fig. \ref{fig:div_mig_break} shows the results of these simulations. The left panel shows the simulation for $m_2< m_{\rm crit}$, or $q>q_{\rm crit}$. We see that that around $\sim 70 R_{\rm g}$ the MMR breaks, as evident from the deviation of the period ratio from the $5:4$ commensurability (middle panel). The resonant angle is plotted at two snapshots of time, at $3\times10^3$ yrs with a small libration, and at $1.03\times 10^5$ yr with full circulation. On the other hand, the right panel shows  the simulation for $m_2> m_{\rm crit}$, or $q<q_{\rm crit}$. The resonance pertains up to very close to the SMBH. The eventual breakup can occur due to differential apsidal precession near $10R_{\rm g}$ if $k$ is low enough, or due to MMR overlap near $R\sim 20-40R_{\rm g}$ for high $k$. We study these conditions in detail in an upcoming paper. 

\section{Evolution in the relativistic regime}\label{sec:Relativistic_evo}

\begin{figure*}
    \includegraphics[width=0.92\textwidth]{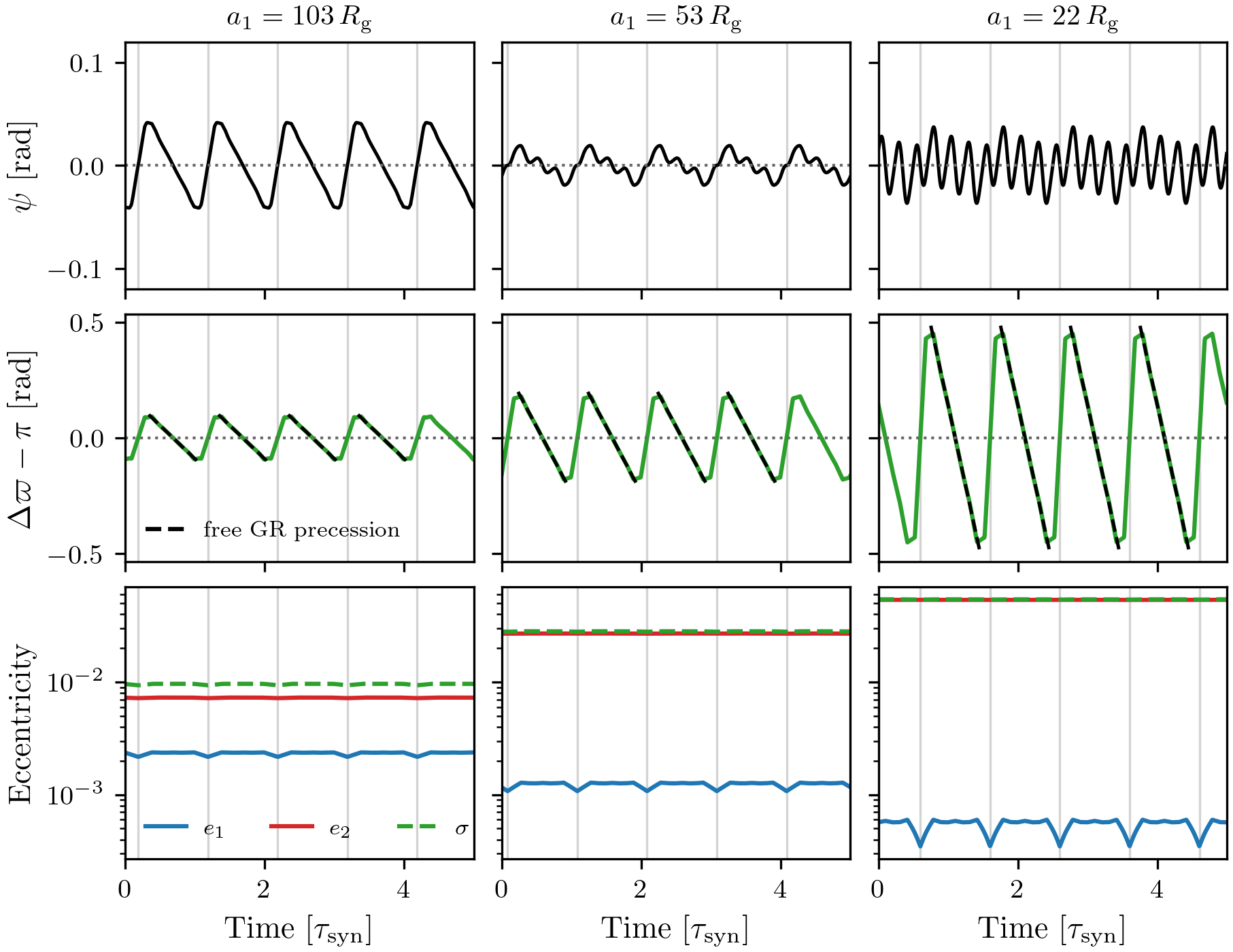}
    \caption{Evolution of the stable $k=5$ MMR system with $m_1=2.59 M_{\odot}$ $m_2=5M_{\odot}$ in the relativistic regime. Each column shows the resonance angle $\psi$, the difference in argument of pericenter $\Delta \varpi$ between bodies $m_1-m_2$, and the eccentricities over a short timescale (five synodic periods) as the system migrates closer toward the SMBH. Notice that the GR induced differential apsidal precession increase $\Delta \varpi$ between synodic periods. This leads to a repartition of eccentricity to the outer body $m_2$, with $e_1\rightarrow0$ as the system approaches the SMBH, while $e_2$ increases.}
    \label{fig:MMR_portrait_numerical}
\end{figure*}

Consider the $2.59M_{\odot}-5M_{\odot}$ MMR system that remains stable in the GW dominated regime. As the system gets closer towards the SMBH, relativistic effects begin to dominate the evolution of the MMR. At first post-Newtonian order, the longitude of pericenter $\varpi$ precesses at rate:
\begin{equation}\label{eq:1PN_precession}
    \dot \varpi_{i, \rm 1PN} = \frac{3(GM_{\rm SMBH})^{3/2}}{a_i^{5/2} c^2}=3\sqrt\frac{GM_{\rm SMBH}}{a_i^3}\frac{R_g}{a_i}.
\end{equation}
Between orbital conjunctions, the orbits therefore differentially precess by an amount:
\begin{equation}
    \Delta \varpi\approx (\dot \varpi_{1, \rm 1PN}-\dot \varpi_{2, \rm 1PN}) P_{\rm syn} = \frac{6\pi (k-1)R_g}{a_2} \left( \zeta^{-5/2} - 1 \right),
\end{equation}
where $\zeta=a_1/a_2 = ((k-1)/k)^{2/3}$ at exact resonance, and  $P_{\rm syn}=(k-1)P_2=2\pi (k-1) \sqrt{a_2^3/GM_{\rm SMBH}}$. 

This differential apsidal precession induced by 1PN terms acts as a secular forcing on the MMR system, modifying the orbital parameters from a standard Newtonian MMR. We observe this numerically in Figure \ref{fig:MMR_portrait_numerical}, which shows the evolution of the orbital angles and eccentricity of the $2.59M_{\odot}-5M_{\odot}$ MMR. The three columns show the resonance angles and eccentricities at decreasing distances from the central supermassive black hole, $a_1 \in \{103R_g, 53R_g,22R_g\}$.

We see through the libration of the resonance angle $\psi$ that the system remains in resonance at all radii (including the highly relativistic region at $a_1=20R_g$), and becoming dominated by the 1PN precession $\Delta \varpi_{\rm GR}$ between conjunctions. The resonance is maintained by the orbital interaction between $m_1-m_2$ near conjunction, where gravitational forces leading up to conjunction anti-align the arguments of pericentre to $\Delta \varpi=\pi$, and $\Delta \varpi$ is less than $\pi$ after conjunction. The balance of $\Delta\varpi_{\rm conj}<0$ during conjunction and $\Delta \varpi_{\rm GR}>0$ from differential apsidal precession between conjunctions cancel over a synodic period, leading to a stable MMR about $\Delta \varpi=\pi$.

The asymmetry between angular momentum exchange between $m_1$--$m_2$ before and after conjunction leads to a small net boost in the outer eccentricity $e_2$, and a decrease in eccentricity $e_1$,
which becomes significant over the EMRI migration timescale. This can be observed in the bottom row of Figure \ref{fig:MMR_portrait_numerical}, showing the bodies individual eccentricities, and combined effective two body eccentricity $\sigma=\sqrt{e_1^2+e_2^2-2e_1e_2\cos{\Delta\varpi}}$ \citep{tamayo2025unified}, at various radii. 

A derivation of the outer body eccentricity growth via Hamiltonian perturbation theory will be presented in an upcoming paper (Moncrieff et al., 2026 in prep). Here we simply note that this leads to an observable EMRI eccentricity for the outer body (see the next section), which distinguishes mEMRIs from isolated disc driven ``wet'' EMRIs \citep{sun2025emriform}.

\textit{Stability:} The resonance angle $\psi=k\lambda_2-(k-1)\lambda_1-\phi_1$ librates in a stable MMR. In the limit where $e_2 \gg e_1$ (which is the case for the relativistic regime),  $\phi_1\approx\varpi_2$, and we reduce the resonance to an exterior one. Setting $\dot \psi=k\dot \lambda_2-(k-1)\dot \lambda_1-\dot \varpi_2=0$ gives a condition for a stable MMR. 
This condition becomes more transparent when viewing (in the low eccentricity limit) the periapsis advance $\dot \varpi_{\rm GR}$ as a difference of radial and azimuthal frequency \citep{yang2019relativistic,sun2025emriform}:

\begin{equation}\label{eq:1PN_freqs}
    \dot \varpi_{\rm GR} =  \omega_{\phi} -  \omega_{\rm r} \equiv2\pi(f_{\rm \phi}-f_{\rm r}),
\end{equation}
where we have introduced the radial (or Keplerian) frequency $f_{\rm r}$ and azimuthal frequency $f_{\rm \phi} = f_r(1+3R/R_{\rm g})$ \citep[to 1PN order, e.g.,][]{Damour1985GRbinary}, and $R$ is the orbital distance to the central SMBH. The change in the mean longitude $\dot{\lambda} = \Omega_k + \dot{\varpi}=f_\phi$ is given by the azimuthal frequency. $\Omega_k=(GM_{\rm SMBH}/a^3)^{1/2}$ is the usual Keplerian frequency.

\begin{figure}
    \centering
    \includegraphics[width=0.45\textwidth]{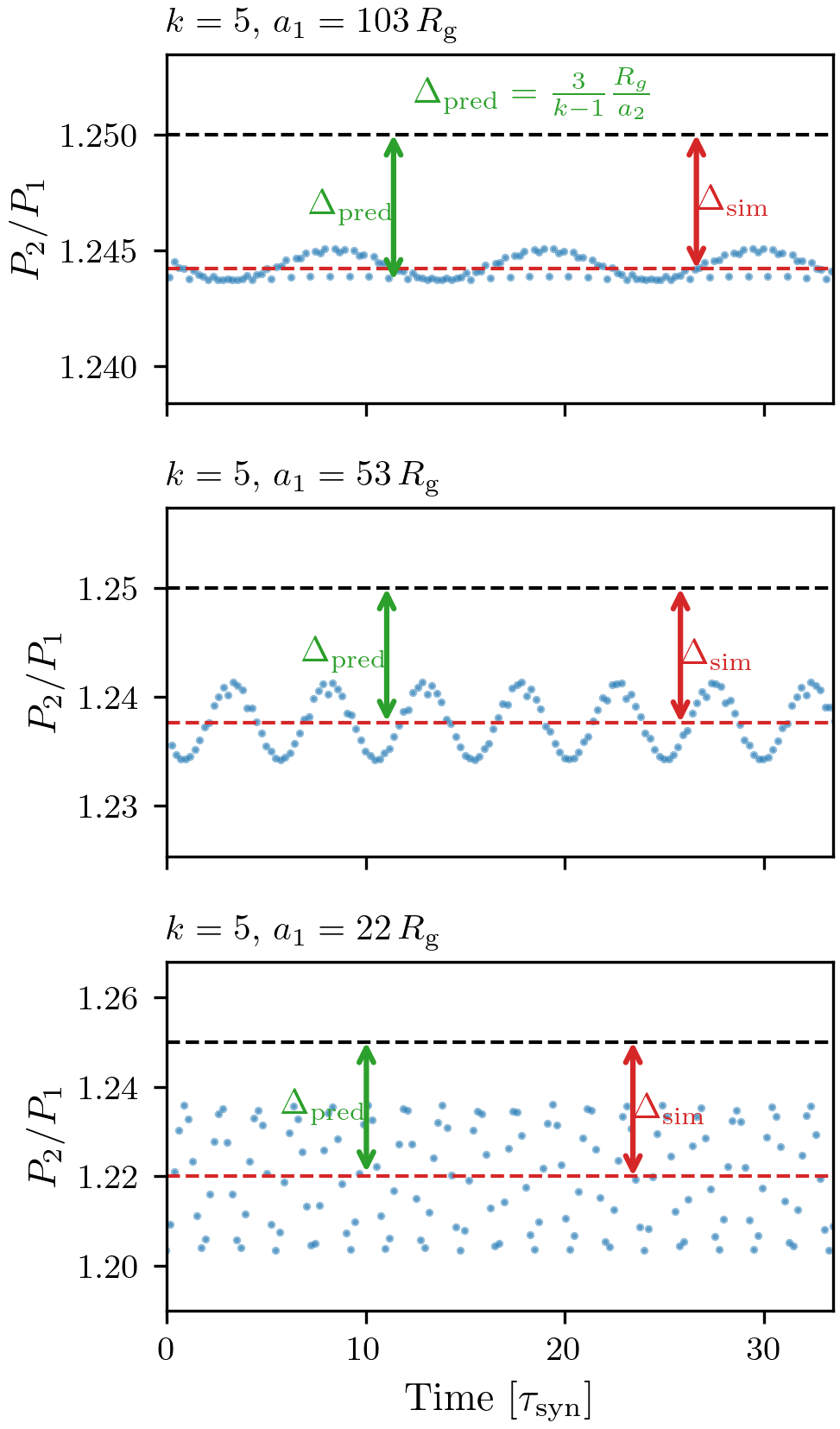}
    \caption{Evolution of MMR period ratio as the 1PN differential apsidal precession between $m_1$ and $m_2$ increases closer to the SMBH. We se that the system is the same as in Figure \ref{fig:MMR_portrait_numerical}. We see that the average offset $\Delta_{\rm sim}=k/(k-1)-P_2/P_1$ increases closer to the central SMBH, matching the analytic prediction $\Delta_{\rm pred}$ very well.}
    \label{fig:MMR_offset}
\end{figure}

\textit{MMR offset:} Averaging over a synodic period, a stable MMR requires:
\begin{equation}
    \langle \dot \psi \rangle_{\rm syn}=(k-1)\langle \dot \lambda_2- \dot \lambda_1 \rangle_{\rm syn}+ \langle\dot \lambda_2-\dot \varpi_2\rangle_{\rm syn}=0.
\end{equation}
 Writing this relation in terms of the synodic frequency, $f_{\rm syn}=\langle \dot \lambda_1  -  \dot\lambda_2\rangle/2\pi$, and using Eq. \ref{eq:1PN_freqs} to write $\dot \varpi_{\rm GR} = \dot \lambda - 2\pi f_{\rm r}$, we obtain the relation:

\begin{equation}\label{eq:fsyn_fr2}
    f_{\rm r, 2} = (k-1) f_{\rm syn}.
\end{equation}
Similar reasoning for the resonance angle $\theta_1=k\lambda_2-(k-1)\lambda_1-\varpi_1=k(\lambda_2-\lambda_1)+(\lambda_1-\varpi_1)$ gives the relation:
\begin{equation}\label{eq:fsyn_fr1}
    f_{\rm r, 1} = k f_{\rm syn}.
\end{equation}
This replaces the Newtonian notion of MMRs, where the orbital periods/frequencies librate about exact period ratios. Instead, at 1PN order, the MMR interaction forces the radial frequencies (which due to periapsis precession are not the same as orbital frequencies) to librate about an integer ratio $f_{r,1}/f_{r,2}=k/(k-1)$. Moreover,
the orbital period ratio is, $P_2/P_1\equiv f_{\phi,1}/f_{\phi,2}<k/(k-1)$ due to 1PN precession. Defining the MMR offset $\Delta\equiv |k/(k-1)-f_{\phi,1}/f_{\phi,2}|$, we observe this offset increasing as the outer MMR semi-major axis $a_2$ decreases - as shown in Fig. \ref{fig:MMR_offset}.

The offset can be predicted analytically by assuming the MMR coupling leads to both $m_1$ and $m_2$ precessing at the angular frequency $\dot \varpi_2$, i.e:
\begin{equation} \label{eq:df}
    f_{\phi,2}-f_{r,2}=\frac{3R_g}{a_2} f_{\phi,2}=f_{\phi,1}-f_{r,1}.
\end{equation}
Note that now we have assumed that the masses $m_1$--$m_2$ are no longer precessing freely, but $m_1$ is forced to precess at the same rate as the more massive $m_2$. We can then compute:

\begin{align}\label{eq:MMR_Delta}
\Delta_{{\rm pred}}	&=\frac{k}{k-1}-\frac{f_{\phi,1}}{f_{\phi,2}}=\frac{f_{r,1}}{f_{r,2}}-\frac{f_{r,1}+3(R_{g}/a_{2})f_{\phi,2}}{f_{r,2}+3(R_{g}/a_{2})f_{\phi,2}} \nonumber \\
	&=\frac{3}{k-1}\frac{R_{g}}{a_{2}}+\mathcal{O}\left(\left(R_{g}/a_{2}\right)^{2}\right),
\end{align}
where we have used  $f_{r,2}(1+3R_g/a_2)=f_{\phi,2}$ and $f_{r,1}/f_{r,2}=k/(k-1)$. This formula, as shown in Fig. \ref{fig:MMR_offset}, agrees very well with the simulated MMR offset, $\Delta_{\rm sim}$. We will see in the next section that these relations manifest in the gravitational wave strain of EMRIs in MMR. 

\section{mEMRI gravitational wave signals}\label{sec:mEMRI_signals}

We consider the gravitational waveform for two EMRIs in the \textit{LISA} band at the same time, migrating via GW emission while locked in a MMR. We obtain the gravitational wave strain $h_+$ and $h_\times$ for each of the sBH $m_1$ and $m_2$, utilising the quadrupole formula for GW radiation from a time-varying mass quadrupole \citep{maggiore08}. The total acceleration entering the quadrupole formula is computed directly within \texttt{TSUNAMI}, following its recent implementation of the corresponding waveforms \citep{gri25lisa}. We then compute the spectral properties and detectability of the GW strain. For all gravitational wave strain computations we assume a face-on observer at a distance $D=100\rm\ {Mpc}$.

\subsection{mEMRI signal observability}

In order to determine the observability of the mEMRI signal,
we take a representative simulation consisting of a $m_1=2.59M_{\odot}$\footnote{This fiducial mass, which lies in the lower mass gap, was chosen to match the secondary component of the highly unequal mass GW190814 event \cite{abbott2020gw190814}. However, the qualitative dynamics of MMRs we study throughout are not impacted by this choice, with the mass ratios $q=m_1/m_2$ and $m_{12}/M_{\rm SMBH}$ determining the dynamics.} 
inner BH in a $k=5$ MMR with a $m_2=8M_{\odot}$ BH orbiting a supermassive back hole of mass $M_{\rm SMBH}=1 \times 10^5 M_{\odot}$. For this mass ratio, the MMR is expected to survive the GW-dominated migration and lead to a mEMRI. The additional AGN disc parameters are $\dot{m}=0.5, \alpha=0.01$, but since gas torques in this regime are negligible compared to the GW back-reaction, it will not influence our results.

\begin{figure*}
    \centering
    \includegraphics[width=0.9\textwidth]{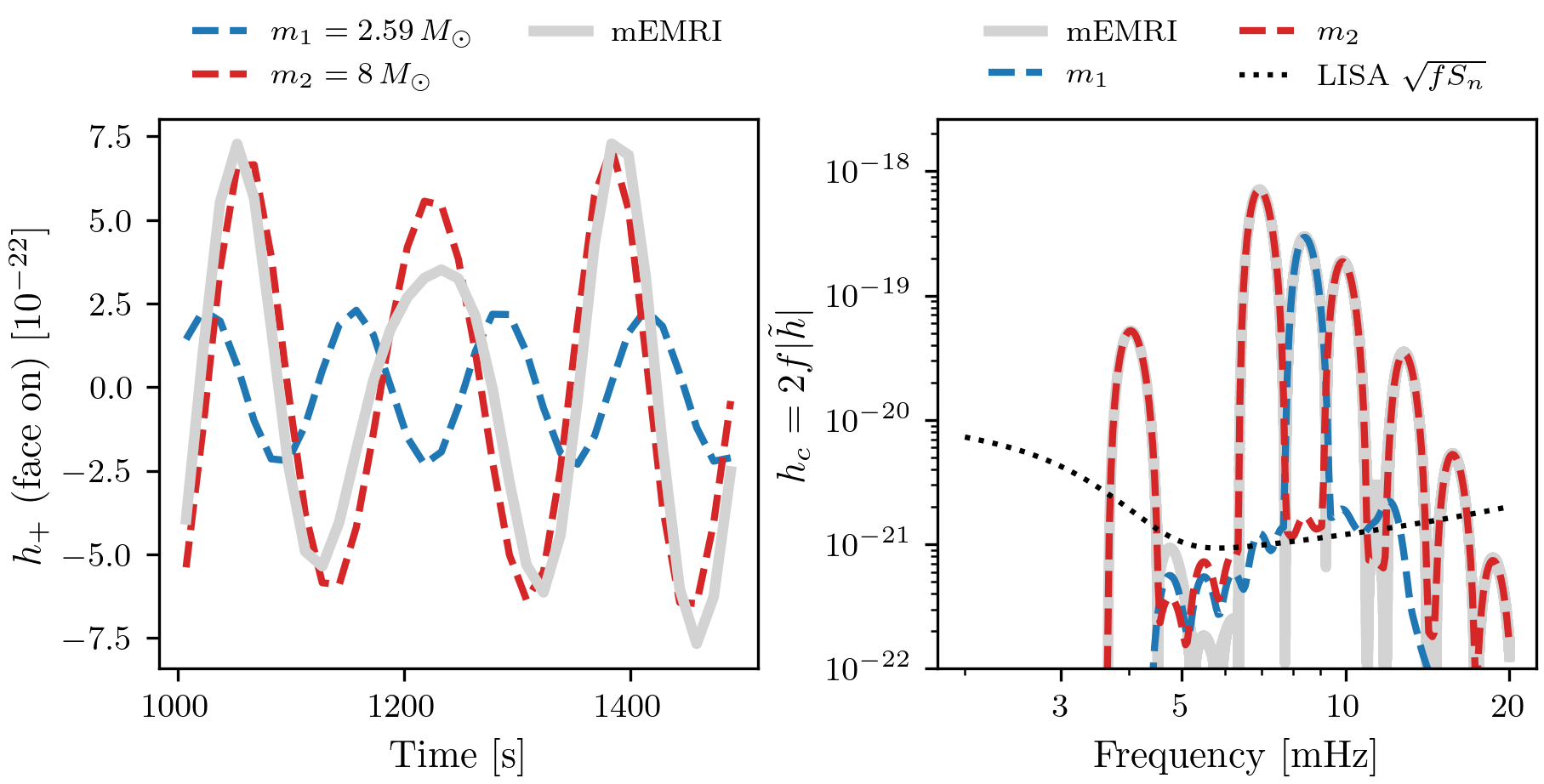}
    \caption{The gravitational wave strain of a mEMRI system with $m_1, m_2=2.59M_{\odot}, 8.00M_{\odot}$ in a $k=5$ MMR situated $a_1 \approx 20R_{\rm g}$ from a central SMBH of mass $M_{\rm SMBH}=1 \times 10^5M_{\odot}$. The left hand figure shows the $h_+$ polarisation gravitational wave strain from the orbit of $m_1$ and $m_2$ over a 500 second period. The right plot shows the characteristic strain $h_c(f)$ for the total gravitational wave strain from the mEMRI system integrated over a time frame of $T=0.5$ years. Note that both the strain from $m_1$ and $m_2$ (shown in blue and red respectively) are well above the \textit{LISA} sensitivity curve.}
    \label{fig:mEMRI_k5_strain}
\end{figure*}

We collect the gravitational wave strain $(h_{+}, h_{\times})$ over an observing time $T=0.5$ years, starting with $a_1\approx 20R_{\rm g}$. The $h_+$ polarisation strain from each mass is shown in a 500 second time segment on the left hand panel of \ref{fig:mEMRI_k5_strain}. To determine the observability of this mEMRI system, we take the one-sided 1D discrete Fourier transform of the strains (as implemented in \texttt{numpy.fft.rfft}):
\begin{equation}
    \tilde{h}_{+,\times}(f)=\frac{\Delta t}{\sqrt{\langle w^2\rangle}} \sum_n h_{+,\times}(t_n) \exp\left(  -2\pi i ft_n  \right) w(t_n).
\end{equation}
where $t_n$ are the uniformly sampled times at which the strain is computed, and $\Delta t$ is the time-step between samples. $w$ is a Hann window implemented to reduce spectral leakage and prevent unphysical strain-frequency peaks of the Discrete Fourier transform.
The detectability of a given signal is determined by comparing the resulting characteristic strain:
\begin{equation}
    h_{\rm c}(f)=2f \sqrt{|\tilde{h}_+(f)|^2 + |\tilde{h}_\times(f)|^2},
\end{equation}
with $\sqrt{f S_n(f)}$, where $S_n(f)$ is the sky and polarisation averaged one-sided effective noise spectral density of the \textit{LISA} detector - which we approximate from the analytic expression given by \cite{Robson2019LISA}. This strain is shown for our system on the right hand panel of Fig. \ref{fig:mEMRI_k5_strain}, where we see that this mEMRI is clearly observable by \textit{LISA}. To characterise this further, we can compute the signal to noise ratio:
\begin{equation}
    \rho^2 = 4 \int \frac{|\tilde{h}_+(f)|^2+|\tilde{h}_{\times}(f)|^2}{S_n(f)} df = \int \left( \frac{h_c(f_i)}{\sqrt{f_i S_n(f_i)}}\right)^2 d \log(f_i). 
\end{equation}

Using only the strain from $m_1$ we get a SNR of 116, while the $m_2$ strain has a SNR of 320. This makes each individual EMRI detectable given a correct waveform for the EMRI systems. The total SNR if both $m_1$ and $m_2$ strains are used is 341. 

We have made several arbitrary choices in our representative setup that can be varied to increase/decrease the SNR of a mEMRI system. SNR scales strongly with observation window $T_{\rm obs}$, observation distance $D$, and inclination angle $\iota$, with \citep{Robson2019LISA}:
\begin{equation}
    \rho \propto \frac{\sqrt{T_{\rm obs}}}{D} \sqrt{1+6\cos^2{\iota}+\cos^4{\iota}}.
\end{equation}
Further analysis of mEMRI detectability is beyond the scope of this paper, and from here on we will focus on the qualitative aspects of mEMRI waveforms that differ from isolated EMRIs.

\subsection{Fundamental frequencies of a mEMRI signal}

Consider two stellar mass BH orbiting a SMBH. Several fundamental frequencies are expected in the strain-frequency diagram. For isolated systems with low eccentricity, the fundamental azimuthal frequencies with the strongest peaks are $2f_{\phi,i}$ for $i=\{1,2\}$, corresponding to varying quadrupole moment of $m_i$ over one orbital period. For isolated eccentric orbits, additional peaks will appear at $f^{(n)}_{e,i}\equiv2f_{\phi,i}+nf_{r.i}$ where $n\in\mathbb{Z}$ \citep[e.g.,][]{Drasco2006EMRIsnapshot}. These additional eccentric lines are a result of the epicyclic motion of $m_i$ about the guiding orbital centre, and are suppressed by an order of $\mathcal{O}(e^{|n|}_{i})$.

For a system in MMR with index $k$, we expect additional frequency peaks $f^{(j)}_{s,2}\equiv2f_{\phi,2}+jf_{{\rm syn}}$ for $j\in\mathbb{Z}$, where $f_{{\rm syn}}=f_{r,1}/k=f_{r,2}/(k-1)$ (see Eq. \ref{eq:fsyn_fr2} and \ref{eq:fsyn_fr1}) is the synodic frequency. These result from the small change in the orbital parameters of $m_1$ and $m_2$ during an orbital conjunction.

\begin{figure*}
    \includegraphics[width=0.87\textwidth]{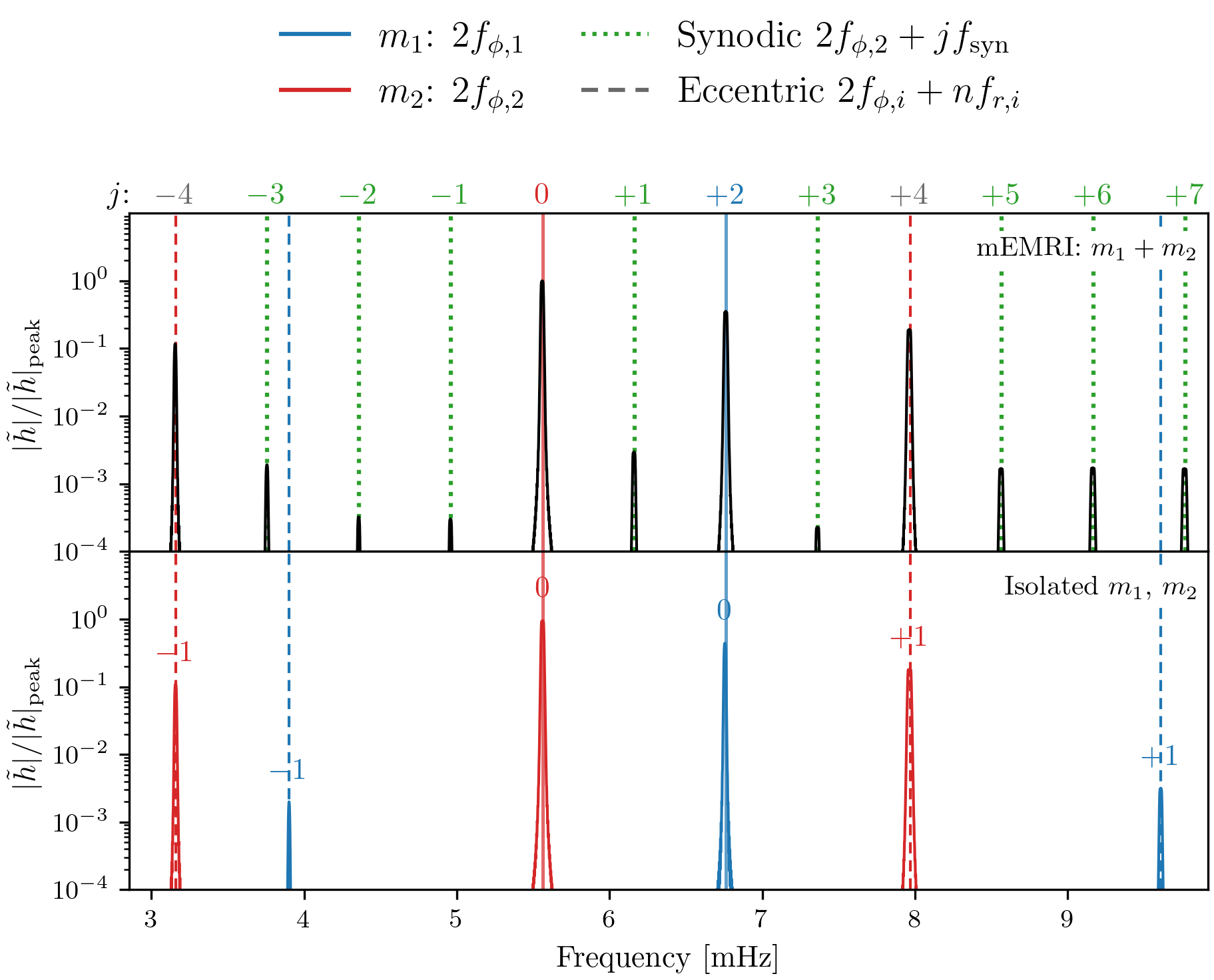}
    \caption{A comparison of the line profile (i.e. short time interval) of the EMRI strain spectra from a mEMRI system (top panel, $m_1$ and $m_2$ in MMR) and reloaded individual EMRI systems with identical initial conditions (bottom, no interaction between $m_1$ and $m_2$). Vertical dashed lines are shown relating the peaks in the spectra relative to the dominant harmonic $2f_{\phi,2}$, which, in the mEMRI case, are spaced by integer multiples of the synodic frequency $f_{\rm syn}=f_{\phi,1}-f_{\phi,2}$.}
    \label{fig:mEMRI_line_comparison}
\end{figure*}

We see that these fundamental frequencies associated with EMRIs via the GW strain in frequency space. Consider again the $k=5$ MMR from Fig. \ref{fig:mEMRI_k5_strain}, but reduce the observation window from $T=0.5$ years to $T=3$ days. We choose this short integration time such that there is negligable frequency evolution of the mEMRI system. We compare the total strain in frequency space $h_+(f)$ for the mEMRI system with the strain of isolated EMRIs with identical initial conditions. The resulting line spectra are compared in Figure \ref{fig:mEMRI_line_comparison}, where the top panel shows the total strain from the mEMRI simulation, while the bottom panel shows in blue (red) the spectra of an isolated EMRI system of mass $m_1$ ($m_2$) initialised around $M_{\rm SMBH}$ with the same orbital parameters of $m_1$ ($m_2$) in the mEMRI simulation.

For the isolated EMRI simulations, we compare the GW strain of the EMRI simulations with the characteristic orbital frequencies, $2f_{\phi,i}+n f_{r,i}$, computed from the N-body orbits (see appendix section \ref{sec:orbital_frequencies} for details). As expected, the dominant peaks in $\tilde{h}_i(f)$ coincide with the eccentric frequencies. Note that in this frequency range only the $n=0,\pm1$ lines are present, and the $n=\pm1$ lines are a factor $\mathcal{O}(e_i)$ weaker than the corresponding $\tilde{h}(2f_{\phi,i})$ peak.

The mEMRI strain contains additional, smaller peaks at integer multiples of the synodic frequency, $2f_{\phi,2}+jf_{\rm syn}$, where $f_{\rm syn}=f_{\phi,1}-f_{\phi,2}$, as expected. Note that all fundamental frequencies in the mEMRI figure are spaced by integer multiples of the synodic frequency. 

\textit{Eccentric and synodic offset:} Interestingly, the $n=\pm1$ lines of the isolated $m_1$ strain are offset slightly from the $j=\pm3$ synodic lines of the mEMRI of Fig. \ref{fig:mEMRI_line_comparison}, while their amplitudes match almost exactly. This offset can be predicted analytically by computing:
\begin{equation}
\frac{\Delta f}{f_{{\rm syn}}} =\frac{\left|f^{(n)}_{e,1}-f^{(j)}_{s,2}\right|}{f_{{\rm syn}}}=\left|nk+2-j+6\frac{R_g}{a_2}\right|,
\end{equation}
where we've assumed that the system is in MMR and used Eq. \ref{eq:fsyn_fr2}-\ref{eq:df} to relate the frequencies to the synodic frequency. The difference $\Delta f$ will be small when $j=nk+2$. For our MMR, $k=5$ and the $j$-values are $j=2\pm5=\{-3,7\}$ for $n=\pm1$. For $k=5$, $\zeta\approx0.862$, plugging in $a_{2}=20R_g/\zeta$ gives $\Delta f/f_{{\rm syn}}=0.258,$ which is in excellent agreement with the value obtained from the numerical simulation in Fig. \ref{fig:mEMRI_line_comparison}. 

\subsection{Impact on EMRI evolution}

\begin{figure*}
    \includegraphics[width=\textwidth]{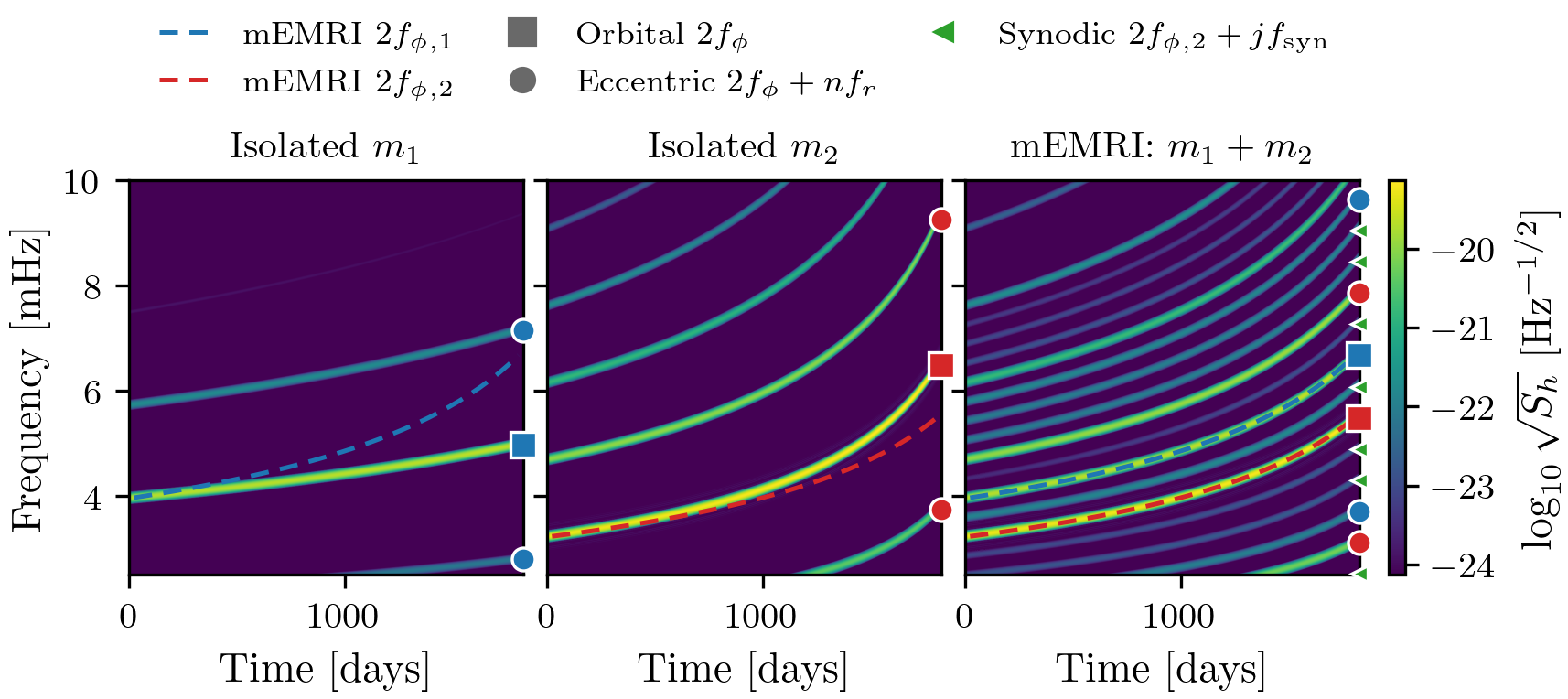}
    \caption{mEMRI frequency evolution (right panel) vs frequency evolution for two individual EMRIs with identical initialised conditions (but not interacting). This system is identical to that shown in Fig. \ref{fig:mEMRI_k5_strain}, except it is initialised with $a_2\approx30R_g$ and migrates to $a_1\approx20R_g$ over $T=5$ years.
    Notice the vastly different chirp of $m_1$ as a lone source in the first panel, with a frequency change much less than the isolated $m_2$ simulation (second panel). In contrast, in the mEMRI $m_1$ migrates in lockstep with the larger $m_2$. Notice also that the combined mEMRI system has faint GW peaks at multiples of the synodic frequency, which are not present in the non-interacting single EMRI systems.}
    \label{fig:mEMRI_freq_comparison}
\end{figure*}

The difference between the gravitational waveforms between isolated EMRIs and mEMRIs becomes stark when we consider the evolution of the orbital frequency peaks. When isolated, the frequency evolution of a (nearly circular) EMRI follows from \cite{Peters64}: 

\begin{equation}
    \frac{\dot f_{\rm GW, i}}{f_{\rm GW, i}} = -\frac{3\dot a_i}{2a_i} = \frac{96G^3M_{\rm SMBH}^2m_i}{5c^5 a_i^4}\propto m_i.
\end{equation}
This formula is derived by setting $\dot E_i=- P_{\mathrm{GW}, i}$, where $E_i=-GMm_i/2a_i$ is the Keplerian orbital energy, and $P_{\mathrm{GW},i}$ is the orbit averaged rate of energy loss from gravitational wave emission in a Keplerian orbit:
\begin{equation}
    P_{\rm GW,i}=\frac{32}{5}\frac{G^4 m_i^2M_{\rm SMBH}^3}{c^5a_i^5} g(e_i),
\end{equation}
where we have approximated $m_i+M_{\rm SMBH}\approx M_{\rm SMBH}$ and defined $g(e_i)=\left( 1+\frac{73}{24}e_i^2+\frac{37}{96}e_i^4  \right)/(1-e_i^2)^{7/2}\approx 1$ for small eccentricity.

When locked in MMR, the ratio of the orbital frequencies must remain constant (up to a small correction from GR, discussed in section \ref{sec:Relativistic_evo}), meaning the frequency evolution of the masses $m_1$ and $m_2$ instead evolve under the combined gravitational wave energy loss of both masses. Setting $a_1/a_2=((k-1)/k)^{2/3}=\zeta$ constant, and solving $\dot E_1 + \dot E_2=- P_{\rm GW, 1}-  P_{\rm GW, 2}$, we get a frequency evolution of the inner EMRI:

\begin{equation}\label{eq:MEMRI_f_evo}
    \frac{\dot f_{\rm GW,1}}{f_{\rm GW,1}} = \frac{96G^3M_{\rm SMBH}^2}{5c^5 a_1^4} \frac{m_1^2g(e_1)+\zeta^5 m_2^2g(e_2)}{m_1+\zeta m_2}=\frac{\dot f_{\rm GW,2}}{f_{\rm GW,2}}
\end{equation}
When $m_2 \gg m_1$, the frequency evolution of the orbital gravitational wave of the inner body $m_1$ will increase by a factor $\zeta^4m_2/m_1 \gg 1$ compared to its evolution in isolation.

We can see this difference in the frequency evolution in the isolated EMRI vs mEMRI case in Figure \ref{fig:mEMRI_freq_comparison}. In this figure, we compare the GW spectrogram of the mEMRI system (integrated for 5 years, with $a_{2,\rm init}=30R_{\rm g}$) with the spectrogram for the reloaded individual EMRI systems. This is similar to the setup considered for Fig. \ref{fig:mEMRI_k5_strain}, but we choose a longer integration time frame and a larger initial radius to observe a greater frequency evolution. This spectrogram is computed using $\texttt{scipy.signal.spectrogram}$\footnote{We used 2048 samples per segment for the spectrogram, with time windows that overlap by 80 per-cent, and applied a Blackman window function.}.

The right panel shows the spectrogram of the total GW strain from the mEMRI integration, showing a red dashed line over the dominant harmonic $2f_{\phi}$ of the body $m_2$, and the $m_1$ line in dashed blue. Notice that the two dominant frequency evolutions of $m_1$ and $m_2$ evolve in lock-step over the five year time-scale, mirroring the joint migration of the MMR pair. Additional frequency lines corresponding to the synodic period modes are also visible.

The two left panels show the frequency evolution of the isolated EMRI systems. Note that while the dominant $m_2$ trajectory has a relatively similar frequency evolution as its corresponding mEMRI track, the $m_1$ system dramatically differs from the isolated case (as expected from Eq. \ref{eq:MEMRI_f_evo}, since $m_2 \gg m_1$). In addition, the synodic lines are absent from the isolated simulations, as expected.

\section{IMPLICATIONS FOR LISA DATA ANALYSIS}\label{sec:LISA_discussion}

\subsection{Degeneracies and environmental effects}

Environmental effects associated with orbital dephasing from gas torques \citep{barausse2007EMRI,kocsis2011emri,Cole:2022yzw}, transient resonances near Kerr SMBHs \citep{Flanagan2012transientResonances,Bonga2019tidalEMRI,gupta2021importance} and/or stochastic migration \citep{Garg2026chaoticMig} can impact EMRI waveforms. Understanding dephasing effects from an external environment is crucial due to the extreme precision required for accurate tests of GR with EMRIs \citep{LISA2025waveform,Arun2022LISAfunPhys}. Here we estimate for the first time the environmental effects on EMRI waveforms associated with the presence of two bodies in MMR. Rewriting Eq. \ref{eq:MEMRI_f_evo} in terms of the gravitational wave frequency $f\approx \sqrt{GM_{\rm SMBH}/a^3}/\pi$, we get:

\begin{align}
    \dot f_1&=\frac{96\pi^{8/3}}{5} \left( \frac{GM_{\rm SMBH}}{c^3}  \right)^{5/3} \frac{\tilde m_1}{M_{\rm SMBH}} f_1^{11/3}\nonumber \\
    &=C\left( \frac{G \tilde{\mathcal{M}_1}}{c^3} \right)^{5/3}f_1^{11/3},
\end{align}\label{eq:chirp_eff}
where we have defined the effective mass:
\begin{equation}\label{eq:m1_eff}
    \tilde m_1 \equiv \frac{m_1^2+\zeta^5 m_2^2}{m_1+\zeta m_2},
\end{equation}
and the effective chirp mass $\tilde{\mathcal{M}_1}\equiv M_{\rm SMBH}^{2/5} \tilde m_1^{3/5}$. The dimensionless constant is $C=96\pi^{8/3}/5$. Similarly, $f_1/f_2\approx \zeta^{-3/2}$, giving:
\begin{align}
    \dot f_2=C\left( \frac{G \tilde{\mathcal{M}_2}}{c^3} \right)^{5/3}f_2^{11/3},
\end{align}
where $\tilde{\mathcal{M}}_2=M_{\rm SMBH}^{2/5} \tilde{m}_2^{3/5}$ and $\tilde{m}_2=\tilde{m}_1/\zeta^4$.

We see that, to first order, there is a degeneracy between an 
$m_1-m_2$ mEMRI pair with effective masses $\tilde{m}_1, \tilde{m}_2$, and isolated EMRIs with the same effective masses. 

Given Eq. \ref{eq:q_crit}, we know that $m_1/m_2\leq \zeta^4$ is required for the $m_1-m_2$ MMR to remain stable in the GW dominated regime. Using $m_2/m_1 \geq \zeta^{-4}$, and dividing Eq. \ref{eq:m1_eff} by $m_1$, we get the inequality:
\begin{equation}
    \frac{\tilde{m}_1}{m_1} = \frac{m_1+\zeta^5 m_2 \frac{m_2}{m_1}}{m_1+\zeta m_2} \geq \frac{m_1+\zeta^5 m_2 \zeta^{-4}}{m_1+\zeta m_2}=1,
\end{equation}
showing that the effective mass of the inner sBH $\tilde{m}_1\geq m_1$ when in MMR. A similar estimate will lead to the inequality $\tilde{m}_2\le m_2$. Thus, if the phase evolution alone is used to determine the mass of an inner (outer) EMRI in MMR, the inferred mass $\tilde{m}_1$ ($\tilde{m}_2$) will be systematically less than (greater than) the true mass $m_1$ ($m_2$).

Interestingly, the observed mass ratio $\tilde{q}=\tilde{m}_1/\tilde{m}_2=\zeta^4 \approx ((k-1)/k)^{8/3}$ will have a discrete set of values for a mEMRI event (up to a small shift from relativistic effects, Eq. \ref{eq:MMR_Delta}).

\subsection{Identifying and characterising mEMRIs}\label{sec:ident_char_mEMRIs}
We have shown that in a MMR pair, the source frame effective mass $\tilde m$ of the individual EMRIs will not match the true mass. The detector frame mass changes according to the redshift, with $m_{\rm det}=(1+z)m_{\rm src}$, but both EMRIs will have the same redshift. This gives a very simple method of identifying mEMRIs, namely, search for EMRIs with matching frequency evolutions $\dot f/f$, and sky localisation posteriors. Statistical association tests would compare the best fit models of isolated EMRIs with independent extrinsic parameters, to a joint mEMRI evolution model with identical parameters (orientation, distance, redshift, SMBH mass, right ascension, declination, etc). 

Joint mEMRI strain evolution models could break the degeneracy between isolated EMRIs with chirp masses $\tilde{\mathcal{M}}_1,\tilde{\mathcal{M}}_2$ by incorporating the gravitational wave strain amplitude of the individual EMRI ampltudes. To leading quadrupole order, the gravitational wave strain of an EMRI is \citep{Peters1963Matthews,maggiore08}:
\begin{equation}
    h_i \approx 4 \left( \frac{G M_{\rm SMBH}}{Dc^2} \right) \left( \frac{Gm_i}{a_i c^2} \right),
    %\propto m_i f_i^{2/3},
\end{equation}
which does not depend on the effective mass $\tilde m$ when two EMRIs are in MMR. This quantity can then be used to determine the mass ratio $q$ of the mEMRI system, via:
\begin{equation} \label{eq:q_true}
    q = \zeta \frac{h_1}{h_2},
\end{equation}
where $h_1$ ($h_2$) are the amplitudes of the peak harmonic of EMRI $m_1$ ($m_2$). With this mass ratio, we can break the mass degeneracy, allowing the determination of the component masses $m_1-m_2$ assuming the two EMRIs are in MMR. 

The requirement that the MMR is stable (Sec. \ref{sec:Relativistic_evo}) leads to several possible consistency checks on the combined mEMRI waveform, particularly if the eccentricity of the outer (more easily detectable) EMRI partner is measurable (e.g., eq. \ref{eq:fsyn_fr2}). This should allow for more efficient searches of mEMRI partners (if there is a low SNR detection) and strong statistical tests of EMRI association (if isolated EMRI searches, e.g., \citealp{Speri2026singleHarmonic}, find candidate mEMRI pairs). Without performing such tests, the presence of mEMRI pairs could exacerbate the global fit problem of \textit{LISA} \citep{Katz2025GlobalFit}, which is already highly challenging for EMRIs \citep{Chua2022Degeneracy}.

\section{DISCUSSION}\label{sec:astro_implications}

\subsection{mEMRI formation and rates}

Predicting the rate of mEMRIs is challenging, owing to the substantial uncertainties in EMRI formation. The fraction of disc-driven EMRIs residing in MMRs depends sensitively on the AGN disc structure and the embedded stellar population. We provide a conservative order of magnitude estimate below. A more robust estimate requires large-scale Monte Carlo simulations across a suite of AGN-disc models \citep[e.g.,][]{mckernan2024mcfacts,vaccaro2026role,Tagawa2026AgnProperties}, combined with prescriptions for MMR capture and breaking \citep{moncrieffmerger2025}. 

We limit ourselves to SMBH masses below $10^6 M_\odot$ for two reasons. First, the peak GW emission $f_{\rm GW}$ must lie in the \textit{LISA} sensitivity band $f_{\rm LISA}$. Taking $f_{\rm GW} \sim \sqrt{GM_{\rm SMBH}/R^3}/\pi$ and $f_{\rm LISA} \sim 10^{-3}\ \rm Hz$ we get a conditions on the SMBH mass: 
\begin{equation}
    M_{\rm SMBH} \lesssim 1\times 10^6M_{\odot} \left( \frac{f_{\rm LISA}}{10^{-3}\ \rm{Hz}} \right)^{-1} \left(  \frac{R}{15R_{\rm g}}\right)^{-3/2}.
\end{equation}
Second, even a modest sBH can open a gap and overcome the migration trap potential for lower mass AGN discs. In particular, the gap mass is $m_{\rm gap}=25M_6\alpha_{-2}^{1/2} \eta_{-2}^{5/2} M_\odot$ where $M_6= M_{\rm SMBH}/10^6M_\odot$, $\alpha_{-2}=\alpha/10^{-2}$ is the normalised $\alpha$-viscosity and $\eta_{-2}=(H/R)/10^{-2}$ is the aspect ratio \citep{kanagawa2018radial}. Thus, even moderate sBH can capture into a MMR and continue migrating to form EMRIs \citep{gilbaum2025escape} and/or mEMRIs.

The total number of host AGNs is the product of the occupation-corrected BH mass function (BHMF) $n_{\rm SMBH}$, the fraction of active galaxies $f_{\rm AGN}$, and the local comoving volume $V_c$. The SMBH number density is estimated via the Schechter-like prescription of \cite{Gallo2019MassFunc}. Integrating their Fig. 2, the local occupation-corrected SMBH number density between $10^{5}-10^{6}M_\odot$ is $n_{\rm SMBH} \approx 2\times 10^{-3}\ \rm Mpc^{-3}$. The AGN fraction in this mass range is around $15-40\%$ \citep{Cho2024_activefrac} based on type 1 AGNs, or $5-22\%$ for the dwarf galaxies estimate of \cite{pacucci21}. We take $f_{\rm AGN}=10\%$ as a conservative estimate, since the \cite{pacucci21} fraction increases with higher SMBH mass. For the local volume ($z\leq0.2$), the distance is $d_L \sim cz/H_0 \approx 850 \ \rm Mpc$ and $V_c= (4 \pi / 3) d_L^3 \approx 2.6 \times 10^9 \rm Mpc^{3}$. Thus, the total number of host AGNs is 
\begin{equation}
    N_\mathrm{hosts} \sim 5 \times 10^5 \times\bigg(\frac{n_\mathrm{SMBH}}{2\times10^{-3}\,\mathrm{Mpc^{-3}}}\bigg)\bigg(\frac{f_\mathrm{AGN}}{0.1}\bigg)\bigg(\frac{V_c}{2.6\times10^{9}\,\mathrm{Mpc^{3}}}\bigg).
\end{equation}

A given AGN may host around $10^3$ sBH \citep[a conservative value compared to other estimates for larger $M_{\rm SMBH}$, e.g.][]{McKernan2018AGNbbhRates}. For a typical disc lifetime of $\sim 10\ \rm Myr$ (and migration time $t_{\rm mig} \le t_{\rm AGN}$) the EMRI rate per AGN is $\mathcal{R}_{\rm AGN}\approx 10^{-4} \ \rm yr^{-1} gal^{-1}$, in agreement up to a factor of a few with the averaged rate of $\sim 10^5\ \rm Gyr^{-1}\ gal^{-1} \approx 10^{-4}  \ \rm yr^{-1} gal^{-1}$ from more detailed Fokker-Planck models of \cite{pan2021formation}. Thus, the total EMRI rate in the local universe is $\mathcal{R}_{\rm EMRI} \sim \mathcal{R}_{\rm AGN} N_{\rm host} \sim 50 \rm\ yr^{-1}$. A fraction $f_{\rm MMR} = f_{\rm cap}\times f_{\rm stable}$ will be a mEMRI in MMR at the \textit{LISA} band. $f_{\rm cap}$ is the fraction of sBH captured to a MMR and $f_{\rm stable}$ is the fraction of MMR pairs that remain stable until they reach the \textit{LISA} band. 

Capture to MMR is expected to be common in the adiabatic limit \citep{batygin2015capture, batygin2023dissipative},
however the efficiency may depend on the radial location and the slope of the torques near the migration trap \citep{Batygin2026overstability}. 
Nevertheless, the adiabatic limit is generally satisfied closer to the SMBH, leading to efficient MMR capture. This can be seen in Figure 12 of \cite{moncrieffmerger2025}, where sBH encounters at the inner migration trap almost always result in MMR capture. We'll take a conservative estimate of $f_{\rm cap}=0.5$. The MMR index $k$ cannot exceed a critical threshold $k_{\rm overlap}$ where Chirikov's resonance overlap \citep{chirikov1979instability} criterion will lead to chaotic motion and escape from the resonance \citep{wisdom1980overlap}
\begin{equation}
    k_{\rm overlap} =0.5 \left( \frac{m_{\rm tot}}{M_{\rm SMBH}}    \right)^{-2/7} \approx 10\left( \frac{m_{\rm tot}}{30M_{\odot}}  \right)^{-2/7} \left( \frac{M_{\rm SMBH}}{10^6M_{\odot}}  \right)^{2/7}.
\end{equation}
The fraction of MMR that remain stable up to the \textit{LISA} band is essentially satisfied by Eq. \ref{eq:q_crit} and will depend on the mass ratio and $k$. For a uniform mass-ratio distribution $q \sim U[0.1,1]$, about $50\%$ of the MMRs remain stable for $k=5$. Lower $k$ values will result in smaller $f_{\rm stable}$. We'll conservatively take $f_{\rm stable} = 0.4$ which corresponds to $k=4$. The overall number of mEMRIs is then 

\begin{equation}
    \mathcal{R}_{{\rm mEMRI}}=10\left(\frac{N_{{\rm host}}}{5\times10^{5}}\right)\left(\frac{\mathcal{R}_{{\rm AGN}}}{10^{-4}\ {\rm yr^{-1}\ gal^{-1}}}\right)\left(\frac{f_{{\rm MMR}}}{0.2}\right)\ {\rm yr^{-1}}.
\end{equation}

\subsection{Comparison to previous studies}
\cite{epstein-mmr2025} analytically studied MMR breaking in AGN discs, attributing differential apsidal precession as the main breaking mechanism close to the SMBH, occurring when the differential precession rate $\Delta \dot{\varpi}_{\rm GR}$ is comparable to the libration rate $2\pi/\tau_{\rm lib}$. In contrast, we find that MMRs remains stable in this regime. As evident in Figure \ref{fig:MMR_portrait_numerical}, the pair develops a phase shift $\Delta \varpi$ during the free precession, which is balanced by apsidal anti-alignment near conjunctions, occuring every synodic time $\tau_{\rm syn} \ll \tau_{\rm lib}$. Hence, the amount of differential precession during the free GR precession phase does not lead to circulation, and the system finds a new equilibrium, analogous to a pendulum in an accelerating frame.

Our condition in Eq. \ref{eq:q_crit} is consistent with the recent studies on MMR formation and break-up in AGN discs. 
\cite{yang2019relativistic} studied the effect of 1PN precession and radiation-reaction on the stability of MMRs, and found that MMRs tend to break once gravitational wave torques overtake gas torques. However, \cite{yang2019relativistic} seem to only consider (nearly) equal mass sBH, which will be destabilised in the gravitational wave dominated regime for all values of $k$ according to Eq. \ref{eq:q_crit}. \cite{peng2023synchronizing} and \cite{peng2024fate} showed that an intermediate-mass-BH and sBH with small mass ratios can survive in MMR down to $10-20R_g$. \cite{sun2025emriform} studies a variety of mass ranges, including a case of mass ratio $10.5/27.9\approx 0.38$ at $k=9$ (their Fig. 4) with a stable mEMRI, and a case of nearly unity mass ratio ($27.5/28.1\approx0.98$) at $k=4$ which was unstable (their Fig. 6). We find all of these results consistent with our criterion in Eq. \ref{eq:q_crit} and conclude that gravitational wave driven MMR breakup depends on the mass ratio and $k$, and not on individual masses, giving a unified picture of MMR breaking observed in previous numerical studies.

\subsection{Caveats and future prospects}

\textit{Accretion disc prescription:}
We use the Sirko-Goodman model of AGN accretion discs \citep{2003MNRAS.341..501S}, which is a modification of the standard Shakura Sunyeav disc model \citep{shakura1973black}. This model breaks down in the highly relativistic regime near the central SMBH, particularly in the regime where the GW emission dominates the gaseous torques. Considerations of relativistic effects in discs have pointed to significantly amplified gaseous torques \citep{Duque2026EMRIdisk}. However, we still expect that once the EMRIs enter the LISA band, gaseous torques will still be subdominant to the migration from GW emission. 

\textit{AGN disc feedback:} sBH can have their own mini-discs that will increase the capture rate and affect  merger timescale \citep[e.g.,][]{whitehead2024}. This is most important once the sBH pair is separated by about their mutual Hill sphere. For our low $k$ MMR, this is not the case and can be safely neglected. The cavity formed by the more massive sBH and/or feedback from the accretion onto the sBH can affect the total torque of the smaller sBH \citep[e.g.,][]{masset2001}, which is not currently modelled.

\textit{Post-Newtonian approximation:} The 3.5PN radiation reaction we use starts becoming innacurate relative to Peters formula for $R \lesssim 30R_g$, and the PN formalism similarly becomes inacurate for $R \lesssim 10R_g$, even for optimal order expansions \citep{Yunes2008PNaccuracy}. To accurately model the evolution of mEMRIs beyond this regime, self-force models are required \citep{LISA2025waveform}. However, we expect that this close to the SMBH, the adiabatic condition will cease to be satisfied as the inner sBH crosses the ISCO. More significant to our results would be the impact of SMBH spin, which we have ignored by assuming a non-spinning SMBH and coplanar sBH orbits. Considering inclined orbits would introduce the additional frequency $f_\theta$ related to the polar angle $\theta$, which would enter into the MMR angle.

\textit{Detection:} In this study we have only simulated the direct gravitational wave strain for such EMRI systems, assessing detectability by comparison to the LISA sensitivity curve. Future work should test the detectability of component mEMRIs in injected LISA noise compared to individual EMRI systems.

\textit{MMR chains:} Long chains of sBH in MMR would likely be less stable than two-body MMRs. The resulting dynamics could be complicated and, to our knowledge, has not been studied. Future mEMRI rate estimates should account for this possibility, as well as resonance breaking by stochastic disc torques \citep{secunda19,trani2025turbulent,Garg2026chaoticMig} or stellar flyby's from the Nuclear star cluster \citep{epstein-mmr2025}.

\section{Summary}\label{sec:summary}
In this paper we have used the few body code \texttt{TSUNAMI} \citep{tsunami-code}, %with 3.5PN corrected equations of motion, 
coupled with an AGN disc model generated with \texttt{pAGN} \citep{pAGN-code} %. that employs analytic prescriptions for type I/II and thermal migration torques
 to study the capture, evolution and stability of MMR from the outer AGN disc to the relativistic regime near the central SMBH. We derived a simple analytic criterion for the stability of these MMRs and analysed the GW signals that arise from such multiple-EMRIs (mEMRIs). Our conclusions can be summarised as follows:

\begin{itemize}
    \item MMRs naturally form in AGN discs due to convergent migration. If the sBH pair overcomes the migration trap, it migrates towards the SMBH (see sketch in Fig. \ref{fig:cartoon}). As GW emission dominates over gas torques, differential migration breaks the resonance if the inner-to-outer mass ratio exceeds the critical value that depends only on the MMR index $k$, $q_{\rm crit}=((k-1)/k)^{8/3}$ (Eqn. \ref{eq:q_crit}).  This yields a simple, unified criterion for MMR survival, encompassing previously studied cases.
    \item For smaller mass ratios, the MMR is stable and the pair migrates as a mEMRI. The orbits will librate around the apsidal anti-alignment equilibrium point $\Delta \varpi = \pi$, even for induced 1PN differential apsidal precession. The outer eccentricity $e_2$ will increase while the inner eccentricity $e_1$ will decrease (see Fig. \ref{fig:MMR_portrait_numerical}).
    \item The relativistic MMR enforces the epicyclic frequencies $f_{r,1}/f_{r,2}=k/(k-1)$ (Eqn. \ref{eq:fsyn_fr1}), while the period ratio drifts from commensurability $f_{\phi,1}/f_{\phi,2}=k/(k-1)-\Delta$, where the offset $\Delta$ grows closer toward the SMBH (see Eq. \ref{eq:MMR_Delta} and Fig. \ref{fig:MMR_offset}).
    \item EMRIs (and mEMRIs) can be observable in the LISA band for $M_{\rm SMBH}\lesssim 10^6 M_\odot$ from $a_2 \sim 15 R_g$. The joint mEMRI GW strain spectral signal contains additional harmonics at multiples of the synodic MMR frequency $f_{\rm syn}=f_{\phi,1}-f_{\phi,2}$, due to periodic interactions of the sBH pair during conjunctions.
    \item The frequency evolution $\dot f/f$ of a mEMRI leads to incorrect effective sBH masses $\tilde{m}_1>m_1$ and $\tilde{m}_2<m_2$. Interestingly, the inferred mass ratio exactly matches to $q_{\rm crit}$ and depends only on the resonance index (while the true mass ratio is $q<q_{\rm crit}$). This degeneracy can be broken if both EMRIs are identified and associated (sec. \ref{sec:ident_char_mEMRIs}), and the true masses can be inferred from the strain amplitudes (Eq. \ref{eq:q_true}). 
\end{itemize}

Future work should provide detailed template banks for mEMRI evolution, devise more thorough statistical tests for the isolated vs mEMRI association, give further consideration to MMR breaking in the strongly relativistic regime, and perform population studies to estimate how the fraction of disc driven EMRIs that are in a MMR scales as a function of the AGN disc properties.

\label{sec:conclusion}

\section*{Acknowledgements}
We thank Marguerite Epstein-Martin, Juliette Becker, and Joseph Livesey for helpful discussions. We thank Nicholas Stone for helpful discussions, and suggesting the name ``mEMRIs''. JWNM, EG and FHP acknowledge support from the OzGrav Research and Innovation Grant. JWNM is supported by funding from the Australian Government Research Training Program. EG acknowledges support from the ARC Discovery Early Career Research Award (DECRA) DE260101802. FHP is supported by a Forrest Research Foundation fellowship. AAT acknowledges support from Fondecyt Iniciacíon grant 11260895.  The authors are grateful for computational resources provided by the OzSTAR Australian national facility at Swinburne University of Technology. Claude was used to generate Figure \ref{fig:cartoon}, based on initial sketches by the authors, and the output SVG files edited. Details of prompts used can be shared on request.

\appendix

\section{Impact of general relativistic effects on resonance parameters}

\subsection{Measuring eccentricities}\label{sec:eccentricity_measure}

There is no unique, gauge-independent, definition of orbital eccentricity in general relativity, which can cause confusion in gravitational wave data analysis \citep{Shaikh2023eccentricity}. For nearly circular EMRIs, we find that eccentricity measures evaluated from instantaneous positions and velocities become dominated by relativistic corrections. This leads to resonance angles that circulate, even when the system is in MMR. At 1PN order, the Newtonian Runge-Lenz vector, $\mathbf{e}_{\rm RL}=\mathbf{v}\times\mathbf{h}/(GM_{\rm SMBH})-\hat{\mathbf{r}}$ (where $\mathbf{h}$ is the orbital angular momentum vector), acquires an additional correction \citep{will2019circle}:
\begin{equation}
    \mathbf{e}_{\rm RL}\approx e_{\rm geo}(\cos{\varpi},\sin{\varpi})-3\frac{R_g}{a}(\cos{l},\sin{l}),
\end{equation}
where $l$ is the true longitude, and 
\begin{equation}\label{eq:e_phys}
    e_{\rm geo} = \frac{r_a-r_p}{r_a+r_p},
\end{equation}
is the ``geometric eccentricity'', computed from the orbital periapsis $r_p$ and apoapsis $r_a$ radii of the orbit. For $e_{\rm geo} \lesssim3R_g/a$, this Runge-Lenz definition therefore tracks the orbital phase rather than the peri centre.

The 2PN quasi-Keplerian radial eccentricity $e_{\rm r,2PN}$ \cite{Memmesheimer2004PNeccen}, which \texttt{TSUNAMI} reports, matches the geometric eccentricity to lower eccentricities. However, this measure vanishes once $e_{\rm geo}\lesssim 0.6 R_g/a$. Following \cite{yang2019relativistic}, we instead measure the geometric eccentricity $e_{\rm geo}$, directly inffered from the position and velocity outputs $(\mathbf r_i, \mathbf v_i)$ (osculating elements will lead to large errors) as described below.

We compute the longitude $l=\nu+\varpi$ as $\arctan{y/x}$, where the $x$ and $y$ components are found from the 3.5PN corrected positions of the particle at a given time-step. We interpolate $r(t)$ with a cubic Hermite spline built from the radial velocity $\dot{r}=\mathbf{r} \cdot \mathbf{v}/r$, locating pericenters and apocenters as roots of the derivative. The eccentricities and semi-major axis are then reported by the interpolated periapsis and apoapsis radii, with $e$ given by Eq. \ref{eq:e_phys} and $a=\frac{1}{2}(r_a+r_p)$.

The true anomaly $\nu=l-\varpi$ then gives the mean anomaly $M$ via $e$ and Kepler's equation \citep{murray1999solar}. The resonance angles are built from the mean longitude $\lambda=M+\varpi$, as in Eq. \ref{eq:res_angle}.

\begin{figure*}
    \includegraphics[width=0.7\textwidth]{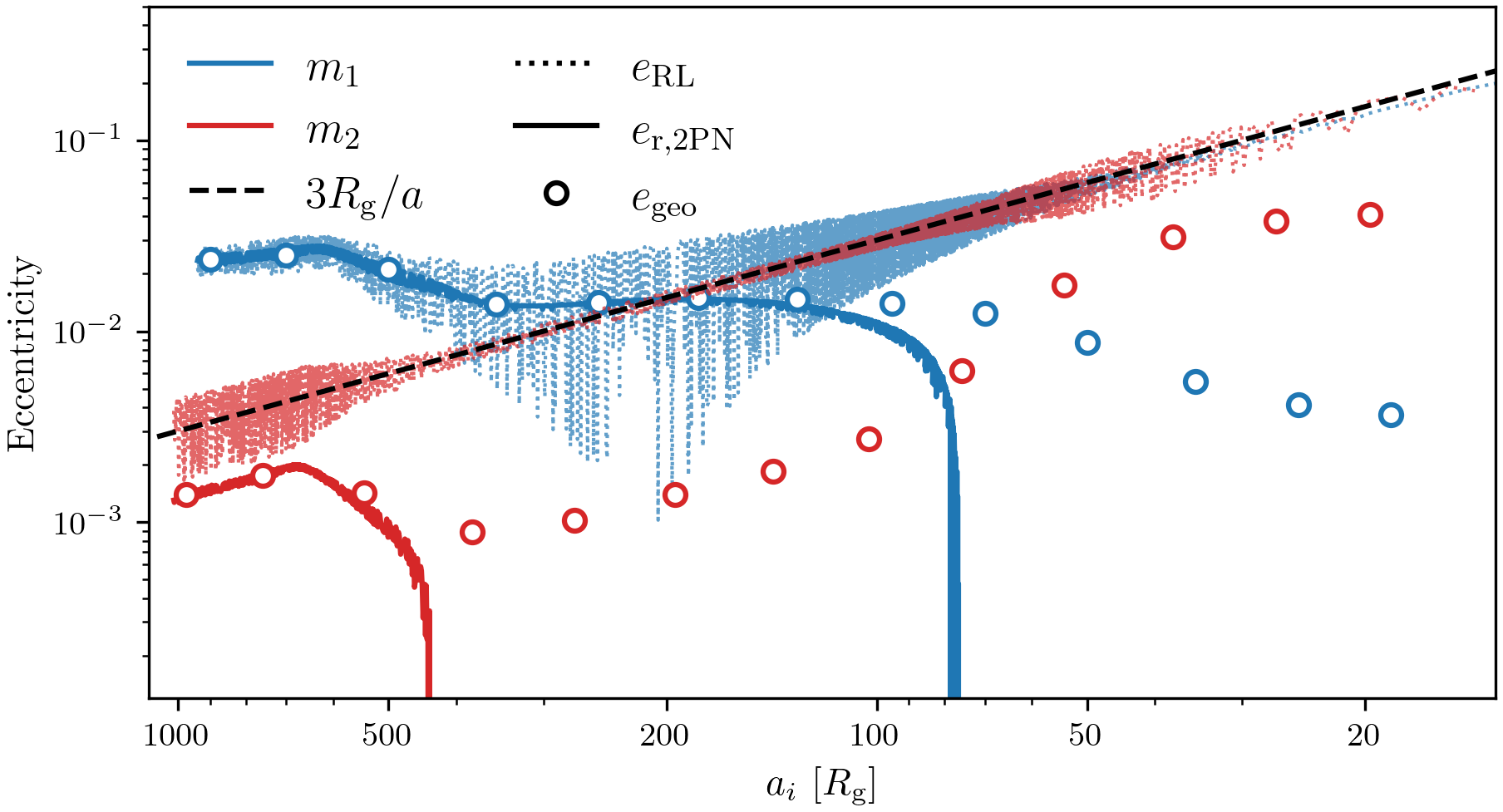}
    \caption{Figure showing the difference between definitions of eccentricity for a simulation with $M_{\rm smbh},m_1,m_2=5\times 10^6M_{\odot},5M_\odot,100M_\odot$ in a $k=9$ MMR. We see that the Runge-Lenz eccentrciity $e_{\rm RL}$ becomes dominated by the 1PN correction when the geometrical eccentricity becomes comparable to $3R_g/a$, while the 1+2PN corrected quasi Keplerian radial eccentricity $e_{\rm r,2PN}$ goes to zero much later, though it vanishes eventually. Meanwhile, the geometric eccentricity $e_{\rm geo}$ remains finite all the way to $20 R_{\rm g}$, and also agrees with both other definitions of eccentricity in the relevant range of validity. Note that the physical eccentricity is only sampled at select times (shown with circular dots) due to the dense output required to accurately compute the physical eccentricity.}
    \label{fig:eccentricity_comparison}
\end{figure*}

\subsection{Eccentricity definition comparison}

This geometric definition of eccentricity $e_{\rm geo}$ and orbital parameters leads to a well defined resonance angle throughout the EMRI evolution, and reduces to the Newtonian value far from the SMBH. Figure \ref{fig:eccentricity_comparison} compares the three measures for a $k=9$ MMR that migrates all the way from $a_2\approx 1000R_g$ to $20R_g$. The AGN disc has parameters $(M_{\rm SMBH}, \dot{m}, \alpha)=(5\times 10^6M_{\odot}, 0.5, 0.01)$, with a pair of masses $m_1, m_2=5M_{\odot}, 100M_{\odot}$. For the inner sBH, all three definitions of eccentricity initially agree down to $a\approx 500R_g$, bellow which $e_{\rm RL}$ becomes dominated by the 1PN correction $3R_g/a$. $e_{\rm r,2PN}$ follows $e_{\rm geo}$ until $\approx100R_g$, though it abruptly drops to zero around $60R_g$, even though the geometrically defined eccentricity remains finite for the whole simulation.  The outer sBH starts with $e_{\rm geo}<3R_g/a$, so $e_{\rm RL}$ is unreliable throughout, and $e_{\rm r,2PN}$ fails at $\sim400R_g$. Both osculating element measures miss the eccentricity growth of the outer body at late times, though in certain simulations (not shown here) this outer eccentricity growth can exceed $\sim R_g/a$, in which case the $e_{\rm r,2PN}$ again becomes an accurate eccentricity measure. We compare the different eccentricity measures across a broader range of orbital configurations and relativistic regimes in a forthcoming paper (Trani et al., in preparation), where we systematically investigate their and domains of applicability, with particular emphasis on the conditions under which osculating-element eccentricities cease to provide a faithful description of the orbital geometry.

\subsection{Orbital frequencies}\label{sec:orbital_frequencies}

Because MMR forcing shifts the orbital frequencies from their isolated values, we measure these frequencies directly from the simulation. The azimuthal frequency $f_{\phi,i}$ is measured as the slope of a linear fit of the unwrapped true longitude $l$ (divided by $2\pi$). This fit is taken over at least one libration period. The radial frequency $f_{r,i}$ is computed from the frequency of the peak Fourier transform amplitude of $r_i(t)$. We limit the possible frequencies to between $0.3f_{\phi,i}$ and $1.6 f_{\phi,i}$ to ignore possible spurious radial frequencies related to gravitational interactions near conjunction. We use a Hann window function for the Fourier transform, as implemented in \texttt{numpy.fft}.

%%%%%%%%%%%%%%%%%%%%%%%%%%%%%%%%%%%%%%%%%%%%%%%%%%
\section*{Data Availability}
The data underlying this article will be provided by the corresponding author upon reasonable request.



\bibliographystyle{mnras}
%\bibliography{example}
% \bibliography{bibtemplate}% if your bibtex file is called example.bib

% Don't change these lines
\bsp	% typesetting comment
\label{lastpage}
\end{document}